\documentclass[conference]{IEEEtran}
\IEEEoverridecommandlockouts

\usepackage{subcaption}  
\usepackage{color}
\usepackage{pifont}
\usepackage{bbm}
\usepackage{stfloats}
\usepackage[short]{optidef}
\usepackage{multirow}
\usepackage{comment}

\usepackage{amsmath,amssymb,amsthm}
\usepackage{ragged2e} 

\usepackage{array}
\IEEEoverridecommandlockouts

\usepackage{cite}
\usepackage{amsmath,amssymb,amsfonts}
\usepackage{algorithmic}
\usepackage{graphicx}
\usepackage{textcomp}
\usepackage{xcolor}
\usepackage{graphicx}

\def\BibTeX{{\rm B\kern-.05em{\sc i\kern-.025em b}\kern-.08em
    T\kern-.1667em\lower.7ex\hbox{E}\kern-.125emX}}
\usepackage{graphicx} 
\begin{document}

\title{Fast and Adaptive Task Management in MEC: A Deep Learning Approach Using Pointer Networks

}
 \author{\IEEEauthorblockN{Arild Yonkeu, Mohammadreza Amini,  Burak Kantarci}\\  \vspace{-0.05in}
 \IEEEauthorblockA{School of Electrical Engineering and Computer Science, University of Ottawa, Ottawa, ON, Canada \\
 \texttt{\{ayonk095, mamini6, burak.kantarci\}@uottawa.ca}}
 \vspace{-0.15in}
 }

\maketitle

\begin{abstract}
Task offloading and scheduling in Mobile Edge Computing (MEC) are vital for meeting the low-latency demands of modern IoT and dynamic task scheduling scenarios. MEC reduces the processing burden on resource-constrained devices by enabling task execution at nearby edge servers. However, efficient task scheduling remains a challenge in dynamic, time-sensitive environments.
Conventional methods—such as heuristic algorithms and mixed-integer programming—suffer from high computational overhead, limiting their real-time applicability. Existing deep learning (DL) approaches offer faster inference but often lack scalability and adaptability to dynamic workloads.
To address these issues, we propose a Pointer Network-based architecture for task scheduling in dynamic edge computing scenarios. Our model is trained on a generated synthetic dataset using genetic algorithms to determine the optimal task ordering. Experimental results show that our model achieves lower drop ratios and waiting times than baseline methods, and a soft sequence accuracy of up to 89.2\%. Our model consistently achieves inference times under 2 seconds across all evaluated task counts, whereas the integer and binary programming approaches require approximately up to 18 seconds and 90 seconds, respectively. It also shows strong generalization across varying scenarios, and adaptability to real-time changes, offering a scalable and efficient solution for edge-based task management.
\end{abstract}

\section{Introduction}


Task offloading and task scheduling in Mobile Edge Computing (MEC) have emerged as critical strategies to meet the ultra-low latency and high reliability requirements of modern Internet of Things (IoT) systems and time-sensitive applications such as autonomous driving, smart manufacturing, and real-time video analytics~\cite{ yang2022deep, jang2023task, palattella2019adhoc}. The proliferation of connected devices, including autonomous vehicles, smart city infrastructure, and industrial IoT components, has led to an exponential growth in data generation, much of which requires immediate processing to enable intelligent decision making and maintain user quality-of-service (QoS). In this context, the traditional cloud computing paradigm, which has centralized resources and introduces high latency due to distant data centers, struggles to meet latency-sensitive requirements due to communication delays and limited scalability in geographically distributed deployments ~\cite{haibeh2022survey, mao2017survey}.

MEC addresses these limitations by decentralizing computation, enabling resource-constrained edge devices to offload intensive tasks to nearby edge servers stationed at the edge of the network. This architecture significantly reduces round-trip latency and removes the processing burden, thus improving responsiveness and extending the battery life of the device ~\cite{ haibeh2022survey, dong2024task}. Efficient task scheduling within MEC environments is particularly important in applications that require high reliability and responsiveness, such as autonomous navigation and real-time data analytics. On-device processing remains impractical for many scenarios due to limitations in energy supply, computational capability, and stringent latency requirements~\cite{yang2023survey}. MEC addresses these challenges by offering powerful computational resources at the network edge~\cite{haibeh2022survey}, allowing the timely execution of tasks. However, effective task management requires adaptive scheduling techniques that can dynamically accommodate diverse and fluctuating workloads~\cite{avan2023state}.

Traditional task scheduling techniques typically lack the flexibility needed to respond efficiently to dynamic MEC environments due to their static nature. Therefore, there is an urgent need for more advanced scheduling algorithms capable of handling the dynamic nature of MEC environments. To overcome these challenges, we propose a novel architecture based on Pointer Networks ~\cite{vinyals2017pointer, vinyals2015order} specifically designed for dynamic and scalable task scheduling in MEC environments. Pointer networks, a class of neural networks used for combinatorial optimization problems, have shown promise in many domains requiring sequence-to-sequence mappings. These networks have shown great generalization capabilities, where they can adapt to various problem instances without retraining. Our model leverages the strengths of pointer networks, allowing effective generalization across diverse scheduling scenarios and rapid adaptation to dynamic conditions. Furthermore, the architecture achieves a low inference time after training, ensuring its practical viability for real-time edge deployment.

\section{Related Works}\label{Sec_related_works}

Conventional MEC task scheduling approaches often rely on heuristic algorithms such as Genetic Algorithms (GA) and Particle Swarm Optimization (PSO), as well as optimization techniques based on Mixed-Integer Linear Programming (MILP) and Mixed-Integer Nonlinear Programming (MINLP)\cite{moshiri2024towards}. While these methods can produce feasible solutions, they typically suffer from high computational complexity due to the NP-hard nature of the problem\cite{wang2020survey, WU2025103754}. This limits their applicability in latency-sensitive scenarios where rapid decision making is crucial~\cite{gong2023dependent}.

To overcome these limitations, recent works have explored the use of machine learning (ML) and deep learning (DL) for task scheduling in MEC environments. Supervised learning models have been proposed to approximate optimal scheduling decisions using historical data~\cite{wang2019deep, mao2017survey}, while reinforcement learning (RL) methods such as Deep Q-Networks (DQN)\cite{luong2019applications}, Actor-Critic methods\cite{zhang2021deep}, and Multi-Agent RL~\cite{xu2021multi} have shown promise in dynamically adapting to changing environments.

Deep reinforcement learning (DRL), in particular, has gained traction due to its ability to learn policies that optimize long-term rewards without relying on labeled datasets. For example, Zeng et al.\cite{zeng2021deep} proposed a DRL-based framework for computation offloading that adapts to network state variations. Similarly, Yang et al.\cite{yang2022deep} developed a DRL-based offloading policy for vehicular networks, demonstrating improved delay performance. However, many of these methods focus more on offloading decisions and less on the scheduling of tasks, which becomes critical in multitask, resource-constrained scenarios.

Other works have adopted sequence modeling approaches, including transformer-based schedulers~\cite{liu2023transformer} and graph neural networks (GNNs) to capture intertask dependencies~\cite{zhou2023gnn}. Although these models improve learning over structured input data, they often incur higher inference latency and are not optimized for real-time execution at the edge.

In the realm of pointer networks, Vinyals et al.~\cite{vinyals2017pointer, vinyals2015order} introduced this architecture for solving combinatorial optimization problems, such as the traveling salesman problem (TSP) and task ordering. However, their use in MEC task scheduling has not been extensively explored. Our work builds on this concept by tailoring pointer networks to the unique requirements of dynamic task scheduling scenarios, including low-latency inference and generalization to unseen task arrival patterns.

In summary, while the existing literature has explored various algorithmic and learning-based strategies for MEC task scheduling, limitations remain in scalability, adaptability, and real-time deployability. This paper addresses these challenges by integrating a pointer-network-based architecture with a novel weighted loss mechanism, optimized for dynamic scheduling in latency-critical edge environments.
Given the sequential nature of task scheduling and deep learning models, we further enhance performance by introducing a weighted loss function that prioritizes early decisions in the scheduling sequence. By assigning greater importance to initial predictions, the proposed framework reduces task drop rates and waiting times as the sequence progresses.

The key contributions of this paper are summarized as follows:
\begin{itemize}
    \item Design a pointer network–based scheduling model capable of real-time inference and generalization across dynamic task scheduling scenarios.

    \item Introduce a weighted loss function that prioritizes early decisions in the scheduling sequence, effectively reducing task drop rates and waiting times when operating in generalized (unseen) environments.
\end{itemize}

The proposed pointer network model achieves a weighted F1 score of 0.88 and a soft accuracy of 0.8917, demonstrating strong generalization across various scheduling scenarios. With inference times consistently under 2 seconds for scheduling 50 tasks, it outperforms integer and binary programming methods, which take 18 and 90 seconds, respectively. These results confirm its suitability for real-time deployment in dynamic edge computing environments. To our knowledge, this is the first learning-based scheduling framework that explicitly considers inference time for practical real-time edge deployment. In the next sections, the details of the proposed method are explained.

The paper is organized as follows: Section \ref{Sec_system_model} introduces the system model, Section \ref{Sec_Formulation} formulates the Joint Scheduling and Task Offloading Problem, Section \ref{Sec_Numerical} presents numerical results, and Section \ref{Sec_Conclusions} concludes with future directions.

\section{System Model}\label{Sec_system_model} 

 Consider a next-generation 5G cellular network environment in which a single gNodeB (gNB) is responsible for providing wireless connectivity and computational support to a set of $N$ connected and automated vehicles (CAVs). The gNB is equipped with a Multi-access Edge Computing (MEC) server, enabling it to perform computational tasks closer to the vehicles, thereby reducing latency and improving the responsiveness of the system. Each CAV generates computationally intensive tasks that cannot be processed locally due to limited onboard resources, and as a result, these tasks are offloaded to the gNB for processing.
\begin{figure}[h]
    \centering 
    \includegraphics[width=.9\linewidth]{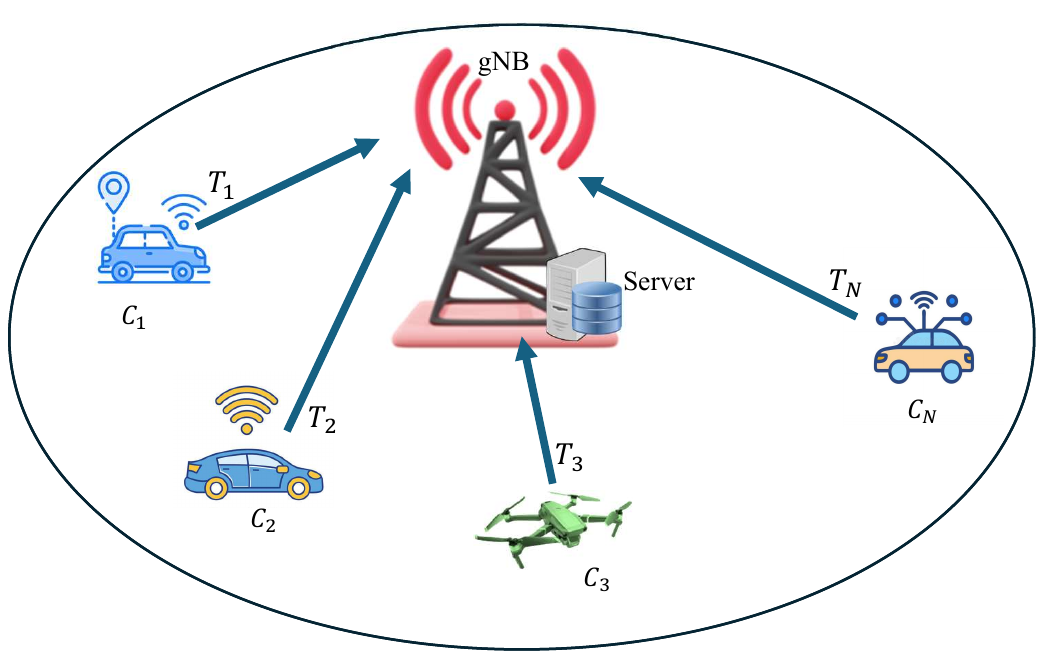}
    \caption{System Model for Task Management  }
    \label{fig:system model} 
\end{figure}
In this setting, CAVs offload their data and task requests to the MEC-enabled gNB via high-speed and low-latency 5G links. These tasks are generally latency-sensitive and require timely execution to support vehicular applications such as autonomous navigation, cooperative perception, or traffic optimization. Efficient task scheduling and resource management at the MEC server become critical to ensure system stability, minimize delay, and satisfy real-time processing constraints.

Let us consider a representative scenario in which several CAVs, denoted as computational vehicles (CVs), transmit their delay-sensitive and resource-intensive tasks to the MEC server. The MEC server must then determine the optimal order of task execution to maximize service efficiency while respecting task deadlines and resource constraints.

Let $\mathcal{C}=\{C_1, C_2, \dots, C_N\}$ denote the set of all connected automated vehicles that submit computational tasks to the MEC server. Each vehicle $C_i \in \mathcal{C}$ sends a task described by a tuple $ {T}_i=\big(t_a^i, t_{exp}^i, t_p^i \big)$.
Here, $t_a^i$ is the arrival time of the task, $t_{exp}^i$ is the task's expiry or deadline time by which it must be completed, and $t_p^i$ is the estimated processing time by the MEC server to execute the task. These parameters are crucial for task scheduling and are assumed to be known or estimated at the time of task arrival. 

This work aims to develop a scalable and adaptive task scheduling mechanism suitable for real-world dynamic environments, where delay-sensitive tasks arrive continuously over time and the execution latency of scheduling algorithms directly impacts system performance. The proposed scheduler must determine an efficient execution order for tasks, ensuring that tasks are completed before their expiry times whenever possible, thus minimizing the drop rate and improving the overall system reliability.

 
\section{Dynamic Task Scheduling- Binary and Integer-based optimization approaches}\label{Sec_Formulation}

Several methods have been proposed for task scheduling and offloading to MEC servers. Common baselines like First-Come-First-Serve (FCFS) and Shortest Task First (STF) are simple and broadly applicable. However, these methods typically fail to optimize resource utilization and critical performance metrics such as task drop ratio, average waiting time, and system responsiveness. In scenarios with high task arrival rates and diverse deadlines, such simple heuristics often lead to suboptimal outcomes, as they do not consider task-specific constraints or optimization as a whole.

A critical issue observed in many existing methods is the disregard for the solver's own execution time. Performance evaluations often focus only on metrics like the drop ratio and waiting time without incorporating the time consumed by the scheduling algorithm itself. This results in misleading evaluations, especially real-time systems, where even a highly optimized task sequence can be ineffective if it takes too long to compute. The oversight becomes significant for workloads sensitive to delays, where scheduling latency directly affects whether a task meets its deadline \cite{gong2023dependent}.

In our work, we explicitly address this shortcoming by incorporating solver execution time into the performance evaluation. We formulate the task scheduling problem using Integer and Binary Programming models that account for solver runtime as a component of the task waiting time. These formulations allow us to capture realistic system dynamics and provide an accurate evaluation of scheduler performance in MEC environments.

Furthermore, in the next section, we propose a deep learning (DL) model based on Pointer Networks that achieves near-optimal task ordering with significantly reduced inference time.

\subsection{Integer Programming Approach}\label{Sec_integer_programming}

To address both scheduling and task offloading, the following optimization problem is formulated. \\
Let $T_i$ be the $i^{th}$ task in the pending queue and $x_j$ be the index of the $j^{th}$ task to be served in the server queue. Furthermore, we denote $s_i$ as the serving order of the $i^{th}$ task, $T_i$. Then

\begin{equation}
    x_j=i \Leftrightarrow   s_i=j
\end{equation}

The waiting time of the task $T_i$ in the server queue can be written based on its order (position) and the processing time of the tasks being served before it. Therefore, the waiting time of the task $T_i$ is written as,

\begin{equation}
    t_w^i=\sum_{j=1}^{s_i-1} t_p^{x_j}
\end{equation}where $t_p^{x_j}$ is the processing time $T_{{x_j}}$. 
The normalized waiting time of $T_i$ is written as,  

\begin{equation}
    t_{w,n}^i=\frac{t_w^i}{\sum_{i^{'}=1}^{N}t_p^{i^{'}}}
\end{equation}

Therefore, the average total waiting time for the tasks is expressed as,  

\begin{equation}
    w_{total}^{ave}=\frac{1}{N} \sum_{i=1}^N t_{w,n}^i 
\end{equation}
The above equation is correct only there is no tasks dropped out of the queue. Since we are dealing with delay-sensitive tasks, some tasks may be dropped if they are expired before they get served. Assuming $d^i$ as the binary index related to the status of the task $T_i$ ($d^i=1$ means task i is dropped), $w_{total}^{ave}$ can we rewritten as,  

\begin{equation}
    w_{total}^{ave}=\frac{1}{N} \sum_{i=1}^N \big(t_{w,n}^i \times (1-d^{i})\big)
\end{equation} in which  

\begin{equation}\label{drop_index}
\begin{split}
    d^i=\begin{cases} 1, \quad \quad t_d^i <t_w^i+t_p^i + t_{exe}^{Int} \\ 
    0 \quad \quad otherwise
    \end{cases}
    \end{split}
\end{equation} where $t_d^i$ and $ t_{exe}^{Int}$ are the deadline of the task and execution time of scheduling $N$ tasks using integer programming. Assuming $t_{cur}$ as the current time of the system at the time the server starts serving the tasks in its queue, then  

\begin{equation}
    t_d^i=t_{exp}^i-t_{cur}
\end{equation} 
Note that $t_{cur}$ reflects the communication delay as well as pending queue waiting time. Accordingly, the drop task ratio can be written as,  

\begin{equation} \label{drop_task_ratio}
    D=\frac{1}{N}\sum_{i=1}^{N}d^i
\end{equation}

 and the optimization problem is written as,

\begin{equation*}
\hspace{-50mm}\underline{\text{\textbf{}}}
\end{equation*}\vspace{-0.3in}
\begin{mini!}|l|[2]                   
    { }{\lambda D + (1-\lambda) w_{total}^{ave} \label{eq:max}}{}{}
	\addConstraint{s_i \in \{1, \cdots, N\}\label{eq:weight}} 
	\addConstraint{s_i \neq s_j \textit{  for  } i \neq j\label{eq:consRel}}
\end{mini!}

\subsection{Binary Programming Approach}\label{Sec_binary_programming}

To formulate task offloading-scheduling as a binary programming, let $X_{i,j} \in \{0,1\}$ be the binary indicator showing the $i^{th}$ task is placed at the $j^{th}$ position in the server queue. $X_{i,j}$ must meet the following constraints. 

\begin{equation}
    \sum_{i=1}^N X_{i,j} \leq 1 \,  \quad  \forall{j}, \quad  \sum_{j=1}^N X_{i,j} \leq 1 \, \quad  \forall{i},
\end{equation}

Then, the position of task $T_i$ in the server, $J_i$ is written as, 

\begin{equation}
    J_i=\sum_{j=1}^N j X_{i,j} \, \quad  \forall{i}
\end{equation} 

Therefore, the waiting time for the $i^{th}$ task in the queue is obtained as, 

\begin{equation}
    t^i_w=\sum_{j=1}^{J_i} \sum_{ i^{'}=1}^N  \, t^{i^{'}}_p X_{i^{'},j}\, \quad  \forall{i}
\end{equation}

Therefore, the normalized average total waiting time considering the dropped tasks is expressed as, 

\begin{equation}\label{waiting_time_2}
    w_{total}^{ave}=\frac{\sum_{i=1}^N \big(t_w^i \times(1- d^i)\big)}{N\sum_{i=1}^{N}t_p^i} 
\end{equation}
where $d^i$ is the drop index for the $i^{th}$ task described as, 

\begin{equation}\label{drop_index}
\begin{split}
    d^i=\begin{cases} 1, \quad \quad t_d^i <t_w^i+t_p^i + t_{exe}^{Bin} \\ 
    0 \quad \quad otherwise
    \end{cases}
    \end{split}
\end{equation}   where $ t_{exe}^{Bin}$ is the execution time of scheduling $N$ tasks using binary programming.

Finally, the binary optimization is formulated as, 

\begin{equation*}
\hspace{-50mm}\underline{\text{\textbf{}}}
\end{equation*}\vspace{-0.3in}
\begin{mini!}|l|[2]                   
    { }{\lambda D + (1-\lambda) w_{total}^{ave} \label{eq:max}}{}{}
	\addConstraint{\sum_{i=1}^N X_{i,j} \leq 1 \, \quad  \forall{j}\label{eq:weight}} 
	\addConstraint{\sum_{j=1}^N X_{i,j} \leq 1 \, \quad  \forall{i}\label{eq:consRel}}
\end{mini!} where $D$ and $w_{total}^{ave}$ are defined in (\ref{drop_task_ratio}) and (\ref{waiting_time_2}), respectively.

\section{Pointer Network-Based Scheduling Approach}\label{Sec_DL_model}

Machine Learning (ML) approaches are gaining traction in task scheduling due to their adaptability to dynamic and scalable environments. Unlike traditional and heuristic methods, which often struggle with fluctuating demands, ML models like pointer networks can learn from historical data and respond quickly to real-time changes, offering greater flexibility and scalability.

Motivated by the dynamic nature of ML approaches, we propose a pointer network-based architecture specifically designed for dynamic and scalable task scheduling. Our architecture uses an encoder-decoder structure without explicit attention layers. The encoder uses an embedding layer to transform the input tokens which represent the individual tasks or start and pad tokens into vector representations. These embeddings are then processed by an LSTM network, which sequentially handles task information to produce hidden and cell states. The decoder mirrors the encoder structure. It uses an embedding layer to decode tasks. 

In the training phase, a teacher forcing strategy is used, where the correct previous task token is provided as input to guide the decoding process. Inference uses sequential decoding, where each step's output token is used for the next step's input. A linear projection layer transforms the hidden states of the decoder into logits corresponding to task-type ids. A masking strategy ensures that only tasks that have yet to be scheduled should be considered. The network also calculates a dynamic mask based on task availability constraints derived from initial task counts, which are updated progressively as tasks are assigned. The exhausting task choices are masked by assigning them a very large negative number (that is, $-10^9$), which excludes them from selection. The pointer network-based architecture accommodates variable-length inputs and outputs through dynamic masking. A masking mechanism is applied based on actual sequence lengths to ensure that irrelevant or padded positions do not influence training or predictions. For training optimization, we use a custom loss function designed specifically for our hybrid pointer network structure. This loss function calculates the average probability assigned by the model to the correct task tokens across sequences, ignoring the padded tokens. To be more precise, it computes the softmax probabilities for each token, gathers the probabilities that correspond to the ground-truth tokens, and then returns the average of these probabilities over valid, nonignored tokens. A sequence that is perfectly predicted yields a loss of zero. 

\begin{figure}[htp]
    \centering
    \includegraphics[width=.8\linewidth]{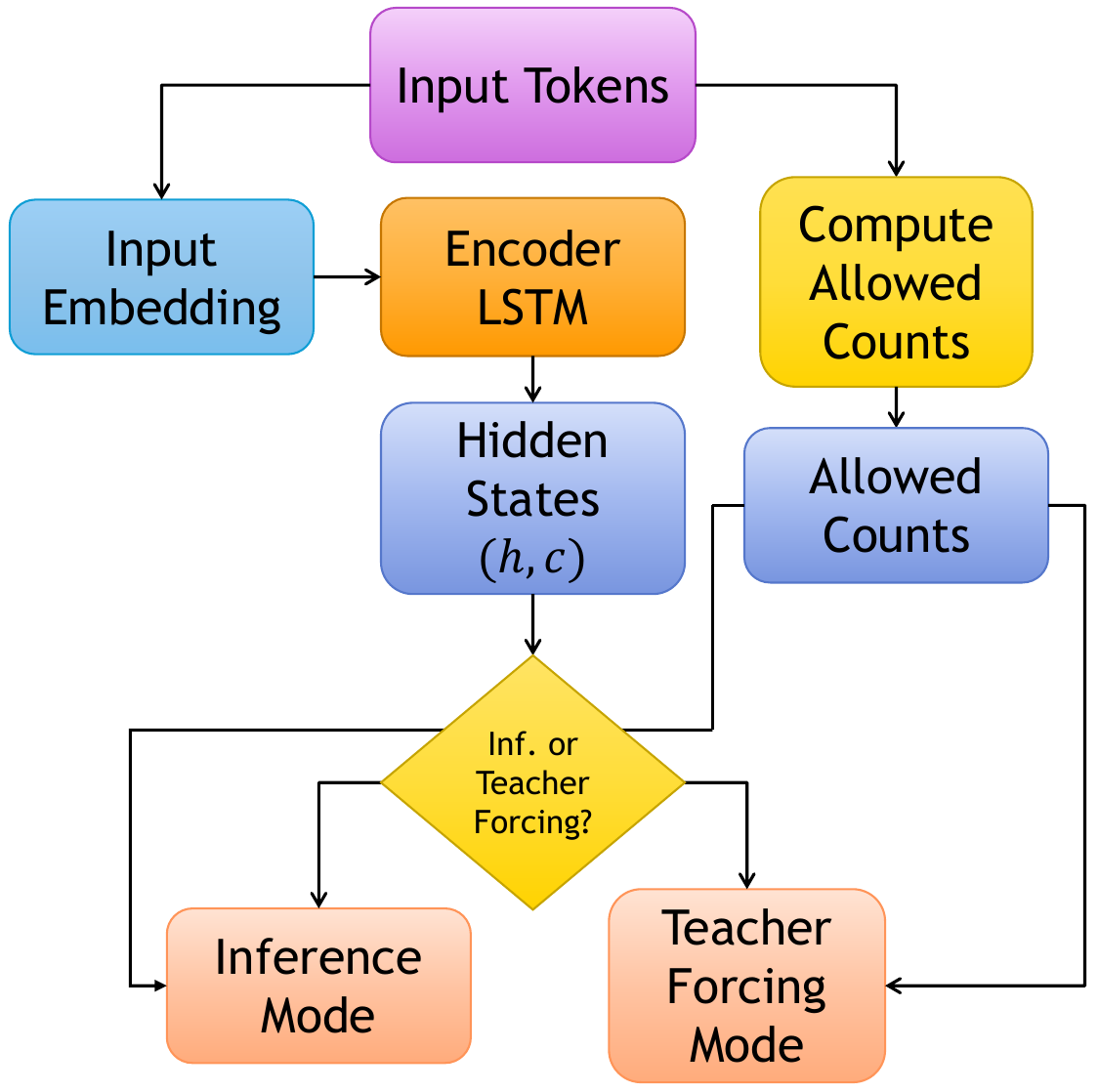}
    \caption{\small Main Encoder Components.} 
    \label{fig:main_encoder_comps}
    
 \end{figure}

\begin{figure}[htp]
    \centering
    \includegraphics[width=.9\linewidth]{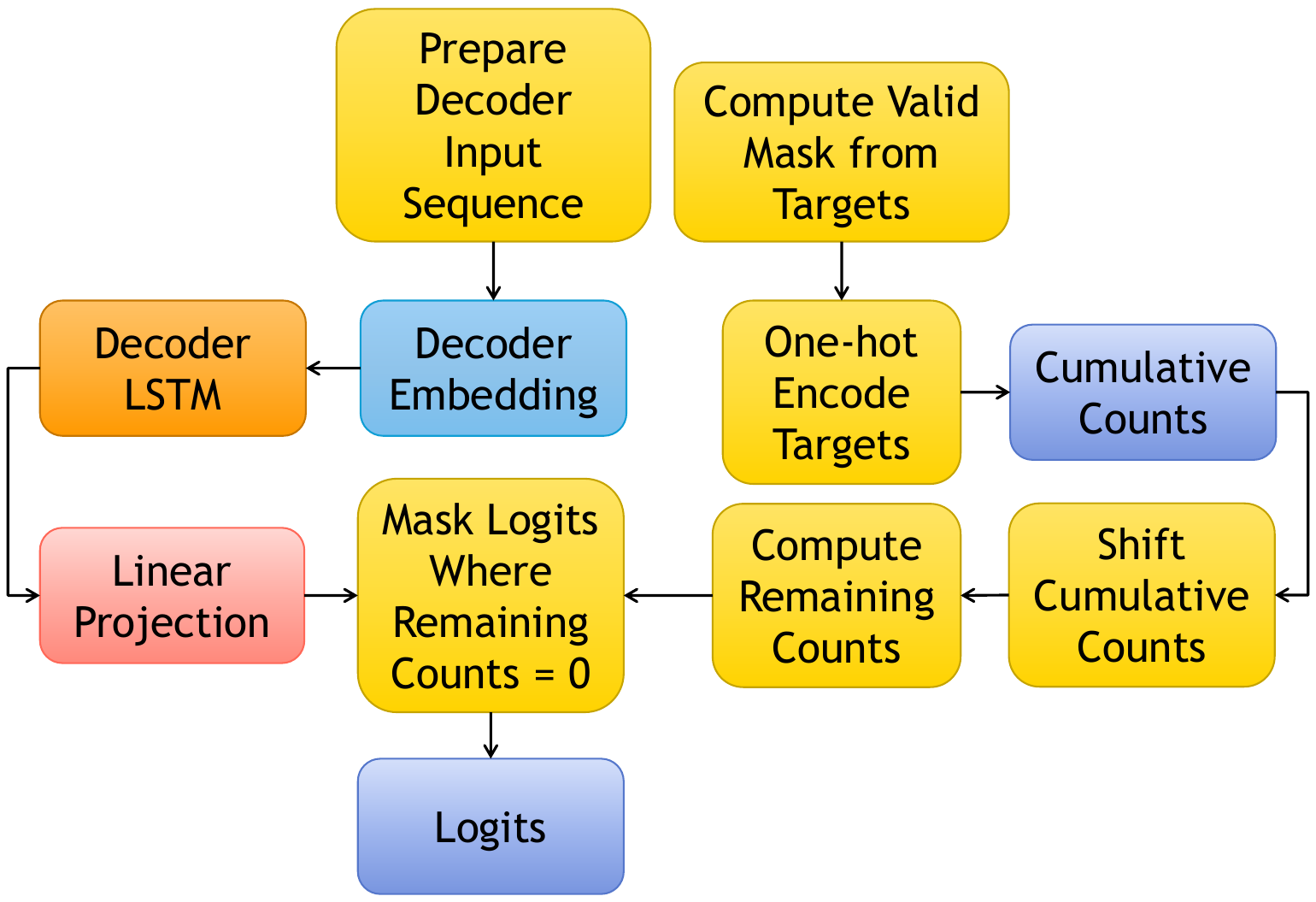}
    \caption{\small Teacher Forcing Mode.} 
    \label{fig:tf_mode}
   
 \end{figure}

\begin{figure}[htp]
    \centering
    \includegraphics[width=.9\linewidth]{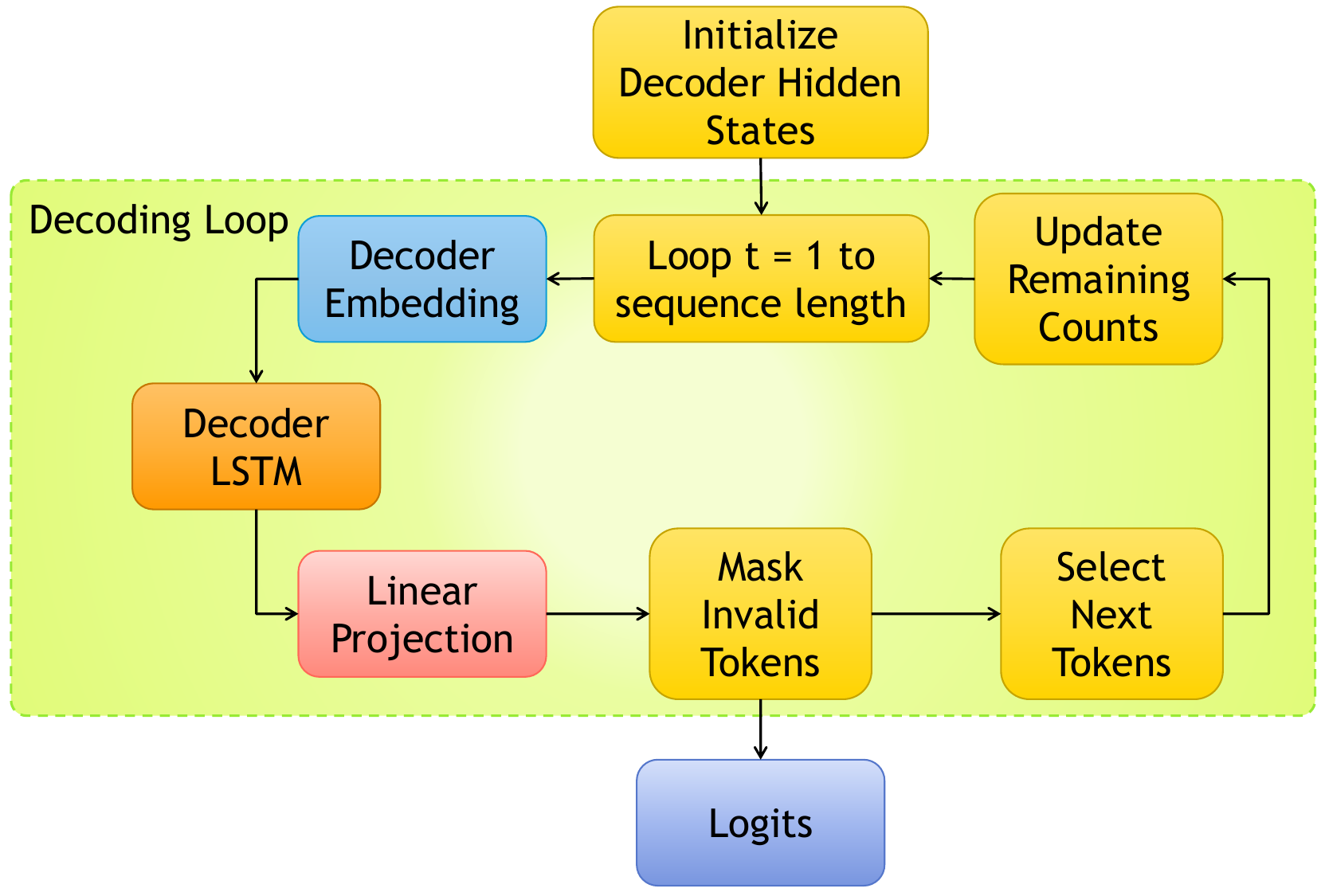}
    \caption{\small Inference Mode.} 
    \label{fig:inf_mode}
    
 \end{figure}

Complementing this, we develop a weighted loss function to improve the model's sensitivity to early predictions in the sequence by applying a decreasing weight vector to token positions. This loss function operates similarly to the previous one, where the weights are applied in the calculation of the average of the probabilities. The weights are also dynamically adjusted on the basis of the sequence length, which normalizes them to a sum of one. However, the weighted loss function is not used for evaluation because it yields the same results as the loss function. The encoder-decoder architecture of our model is depicted in Fig.~\ref{fig:main_encoder_comps}, which shows its components, including the embedding layer and LSTM units.  During training, we opted not to use teacher forcing, as the results showed better generalization and prediction without it. The inference process, depicted in Fig.~\ref{fig:inf_mode} uses sequential decoding, where each output token is fed as input for the next decoding step. Fig.~\ref{fig:tf_mode} shows the teacher forcing mode, which was evaluated but excluded from training due to slightly lower performance.


\begin{table}[htbp]
\caption{Hyperparameter Setting}
\vspace{-3mm}
\begin{center}
\begin{tabular}{|c|c|}
\hline
\textbf{Parameter} & \textbf{Value} \\
\hline\hline
Batch Size & 128\\
\hline
Initial Learning Rate & 0.001 \\
\hline
MaxEpoch & 20 \\
\hline
Hidden Layer Size & 128 \\
\hline
FC-1 Input Layer & 128 \\
\hline
FC-1 Output Layer & 9\\
\hline
Training Optimization Method & Adam\\
\hline
Loss Function & SoftSequenceAccuracyLoss \\
\hline
Training/Test/Validation ratio & 80\% / 10\% / 10\% \\
\hline
\end{tabular} 
\end{center}
\label{tab:hyperparameter}
\end{table}

\section{Numerical Results and Comparison }\label{Sec_Numerical} 
This section describes the process of generating the dataset used for training and evaluating the performance of the proposed architecture.

\subsection{Generating Data Set}
To train our pointer network-based scheduling architecture, we generated a synthetic yet realistic dataset. Initially, tasks were defined by pairs of attributes, processing times, and deadlines. These attribute pairs were then assigned to the task types' identifiers. To be more specific, nine unique task types were defined with fixed processing times and deadlines (i.e., type 0 with attributes $(10, 50)$, type 1 with $(10, 100)$ and so on). Random sequences of these task-type IDs were then generated, allowing repetition within a single sequence to simulate real-world scenarios. With the maximum sequence length of 10, sequences that are shorter than this were zero-padded. A sequence with an actual length of 5 yields $9^5=59049$ possible unique sequences. To identify the optimal task ordering from these randomized inputs, we used a genetic algorithm (GA) to minimize task drop ratios and to reduce waiting times. Each dataset sample consists of three components: a randomly generated initial sequence of task-type identifiers, zero-padded to length 10; the corresponding optimal ordering determined by the GA; and the actual sequence length before padding.

\begin{table}[htbp]
\caption{Parameter Setting for GA Dataset Generation }

\begin{center}
\begin{tabular}{|c|c|}
\hline
\textbf{Parameter} & \textbf{Value} \\
\hline\hline
Population size & 200\\
\hline
Number of generations & 500\\
\hline
Patience & $100$ \\
\hline
Mutation Type & Swap\\
\hline
Mutation Probability & 0.3\\
\hline
Parent Selection Type & Tournament\\
\hline
Elitism Percentage & 5\% \\
\hline
Number of parents mating & 30\% of population size \\
\hline
Gene Space & Permutations of Task Indices \\
\hline
Initial Population & STF, SDF, and Random Permutations \\
\hline
Crossover Type & Ordered Crossover \\
\hline
\end{tabular}  
\end{center}
\label{tab:ga_hyperparameters} 
\end{table}

\subsection{Performance Evaluation}

\begin{figure}[htp]
    \centering 
    \includegraphics[width=1\linewidth]{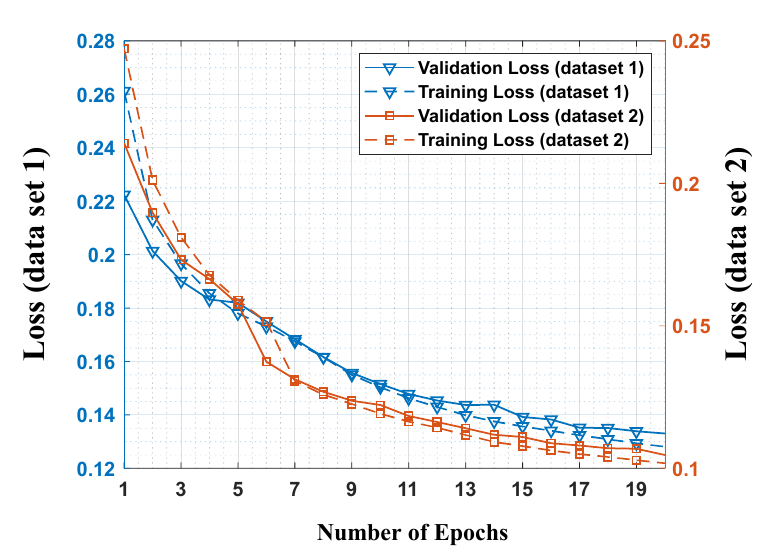}
    \caption{Loss vs. number of epochs}
    \label{fig:performnce} 
\end{figure}

After dataset generation, the proposed network is trained and evaluated using two datasets containing $307,290$ and $566,340$ samples, respectively. Fig. \ref{fig:performnce} presents the training and validation losses across epochs for both datasets. Dataset 2 consistently shows lower loss values, indicating faster convergence and better generalization. Validation losses closely follow training trends, suggesting stable learning. Final validation losses reach $0.1331$ for dataset 1 and $0.1047$ for dataset 2.

We evaluated the performance of the trained pointer network against traditional heuristic-based scheduling methods such as FIFO (First-In-First-Out) and Shortest Deadline First (SDF). Performance metrics considered were task drop ratio and average waiting time. Table~\ref{tab:dataset_features} Set 1 contains $307,290$ samples, while Set 2 is nearly twice as large, with $566,340$ samples. Both datasets were constructed using three possible processing times $[10,20,30]$ and three corresponding deadlines $[50,100,150]$, reflecting diverse yet realistic time-sensitive task conditions encountered in dynamic task scheduling scenarios \cite{3GPP.22.261,wang2020imitation}. The average soft accuracy is a relaxed form of the sequence accuracy where partial correctness is considered. For example, if the correct sequence is $[1,2,3]$ and the models predicts $[1,3,2]$, then the loss is $0.66$, rather then $0$, since one task is in the correct position. The average soft precision and recall account for the precision and recall of partially correct sequences.The average soft accuracy, soft precision, and soft recall can be written as follows: 

\begin{equation}
\text{Average Soft Accuracy} = \frac{1}{M} \sum_{m=1}^{M} \frac{\sum_{i=1}^{L_m} P_{m,i}(y_{m,i})}{L_m}
\end{equation}

\begin{equation}
\text{Average Soft Precision} = \frac{1}{C} \sum_{c=1}^{C} \frac{\text{soft TP}_c}{\text{soft TP}_c + \text{soft FP}_c + \epsilon}
\end{equation}

\begin{equation}
\text{Average Soft Recall} = \frac{1}{C} \sum_{c=1}^{C} \frac{\text{soft TP}_c}{\text{soft TP}_c + \text{soft FN}_c + \epsilon}
\end{equation}

where \(M\) is the number of sequences, \(L_m\) is the number of tokens in sequence \(m\), \(C\) is the number of classes, and \(\epsilon\) is a small constant to avoid division by zero. Here, \(P_{m,i}(y_{m,i})\) denotes the predicted probability assigned to the correct token at position \(i\) in sequence \(m\). Soft true positives (soft TP), soft false positives (soft FP), and soft false negatives (soft FN) are accumulated probability scores across all tokens and classes.

\begin{table}[h!]
\centering
\caption{Dataset Characteristics}
\begin{tabular}{|l|c|c|c|c|}
\hline
\textbf{Dataset} & \textbf{\# Samples} & \textbf{Poss. Processing Times} & \textbf{Poss. Deadlines} \\
\hline
Set 1 & $307290$ & $[10, 20, 30]$ & $[50, 100, 150]$ \\
\hline
Set 2 & $566340$ & $[10, 20, 30]$ & $[50, 100, 150]$ \\
\hline
\end{tabular}

\label{tab:dataset_features}  
\end{table}

Table~\ref{tab:model_metrics} presents the evaluation metrics of our model in both datasets. In particular, the model achieves strong performance in Set 1 and Set 2, with a weighted F1 score of $0.86$ and an average soft accuracy of $0.8790$. In Set 2, the model maintains performance with a slight improvement, reaching a weighted F1 score of $0.88$ and an average soft accuracy of $0.8917$. 

\begin{table}[htbp]
\centering
\caption{Comparison of Model Metrics for Different Datasets}
\resizebox{\columnwidth}{!}{%
\begin{tabular}{|l|c|c|c|c|c|c|c|c|c|}
\hline
\textbf{Dataset} &  \textbf{Weighted F1} & \textbf{Avg. Soft Acc} & \textbf{Avg. Soft Pre} & \textbf{Avg. Soft Rec}  \\
\hline
Set 1 & $0.86$ & $0.8790$ & $0.8636$ & $0.8636$ \\
\hline
Set 2 & $0.88$ & $0.8917$ & $0.8791$ & $0.8791$  \\
\hline
\end{tabular} 
}
\label{tab:model_metrics} 
\end{table} 

\begin{figure}[htp]
    \centering 
    \includegraphics[width=.9\linewidth]{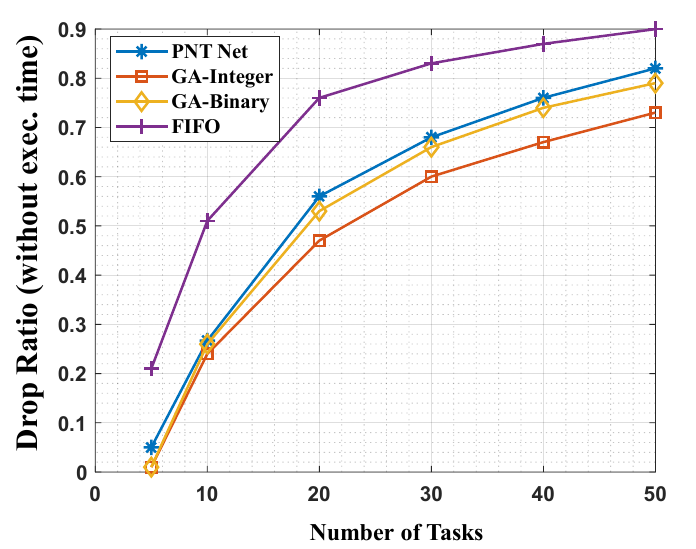}
    \caption{Average drop ratio vs Number of tasks without considering the execution time}
    \label{fig:drop_no_exe} 
\end{figure}

\begin{figure}[htp]
    \centering 
    \includegraphics[width=.9\linewidth]{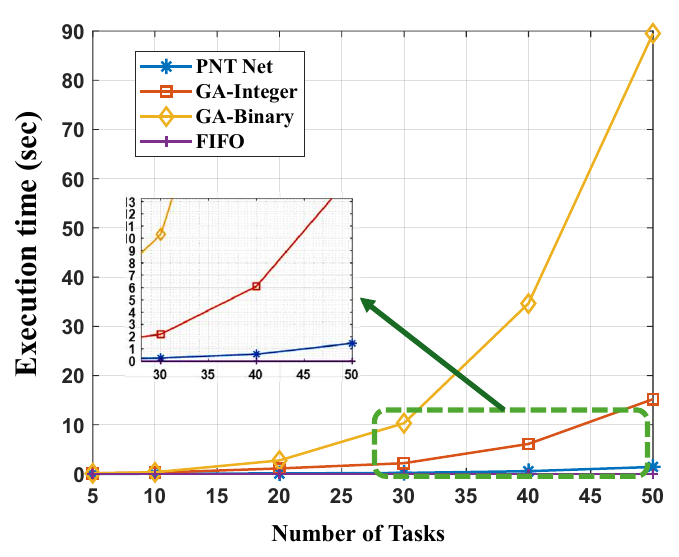}
    \caption{Average execution time vs number of tasks}
    \label{fig:execution} 
\end{figure}

\begin{figure}[htp]
    \centering
    \includegraphics[width=.9\linewidth]{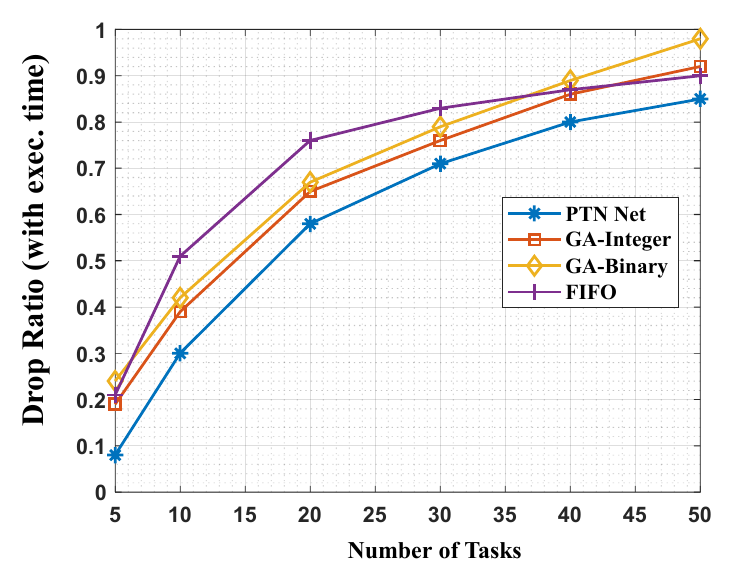}
    \caption{Average drop ratio vs number of tasks after adding the execution time}
    \label{fig:drop_exe} 
\end{figure}

The performance of our model (PNT Net) is evaluated against baselines including FIFO, Integer-based GA (GA-Integer), and Binary-based GA (GA-Binary). Fig. ~\ref{fig:drop_no_exe} presents the average task drop ratio across different numbers of tasks, without considering the execution time of the scheduler. Fig. ~\ref{fig:drop_exe} gives a more realistic perspective by showing the average execution time across the different numbers of tasks of the respective schedulers. Finally Fig. ~\ref{fig:execution} shows the average drop ratios after execution time has been included. This shows the superior scalability and runtime efficiency of the pointer network model compared to GA-based methods, which incur higher computational costs and overhead.

One of the key findings in our evaluation is the trade-off between the quality of the task sequences generated by the Genetic Algorithm (GA) and the algorithm's execution time. While GA yields sequences with near-optimal scheduling accuracy, it comes at a cost: high execution time as the number of tasks increases. In delay-sensitive environments, scheduling time directly contributes to system performance, especially the overall task drop ratio. Our results show that although GA-based schedules are effective in minimizing the drop ratio, their execution time increases the actual drop ratio when it is factored into the system's total response time. Our pointer network model offers near-instantaneous inference, resulting in lower drop ratios and better performance for real-time deployment. By learning from GA-optimized schedules offline, the pointer network can approximate high-quality predictions with drastically reduced latency. For example, the Integer-based GA requires an average execution time of $6$ seconds for $40$ tasks, while our pointer network model requires over $0.5$ seconds, which yields a 12× reduction in execution time.

\section{Future Work} \label{Sec_FutureWorks}
While the proposed pointer network-based scheduler shows high accuracy, low inference latency, and strong generalization, there are several important directions to explore for further improvement and deployment scalability in MEC environments. As part of future work, we plan to extend this framework to multi-agent cooperative scheduling across distributed edge nodes, incorporate meta-learning and reinforcement learning for end-to-end training in dynamic environments, and explore energy-aware scheduling mechanisms to further optimize resource utilization in MEC systems.

\section{Conclusions} \label{Sec_Conclusions} 
In this paper, we addressed the critical challenge of dynamic and efficient task scheduling in Mobile Edge Computing (MEC) environments. We examined the limitations of conventional heuristic and optimization-based approaches, as well as the scalability and adaptability issues associated with existing deep learning models. To overcome these challenges, we proposed a Pointer Network-based deep learning architecture designed for real-time task scheduling in dynamic edge computing scenarios. The proposed model demonstrates strong generalization across varying task patterns, low inference latency, and robust adaptability to fluctuating workloads and network conditions. Experimental evaluations confirm the effectiveness of the model in reducing task drop ratios and improving overall scheduling performance, where our model achieves lower drop ratios and waiting times than baseline methods when execution time is included while operating at less than 2 seconds per task count, and a soft sequence accuracy of up to $89.2\%$.

\section*{Acknowledgment} 

This work was supported in part by funding from the Innovation for Defence Excellence and Security (IDEaS) program from the Department of National Defence (DND) in part by the Natural Science and Engineering Research Council (NSERC) CREATE TRAVERSAL Program, and in part by the Ontario Research Fund-Research Excellence program under grant number ORF-RE012-026.

\bibliographystyle{IEEEtran}

\end{document}